\documentclass[twocolumn]{aa}

\usepackage{graphicx}
\usepackage{txfonts}
\usepackage{natbib}
\bibpunct{(}{)}{;}{a}{}{,} 

\begin{document}

   \title{The ignition of thermonuclear flames in Type Ia supernovae}

   \author{L. Iapichino\inst{1,2}
          \and
          M. Br\"uggen\inst{3}
          \and
          W. Hillebrandt\inst{1}
          \and
          J.C. Niemeyer\inst{2}
          }

   \offprints{L. Iapichino}

   \institute{Max-Planck-Institut f\"ur Astrophysik,
              Karl-Schwarzschild-Str. 1, D-85741 Garching, Germany \\
              \email{wfh@mpa-garching.mpg.de}
              \and 
              Universit\"at W\"urzburg, 
              Am Hubland, D-97074 W\"urzburg, Germany \\ 
              \email{[luigi,niemeyer]@astro.uni-wuerzburg.de}
              \and  
              International University Bremen, 
              Campus Ring 1, D-28759 Bremen, Germany \\
              \email{m.brueggen@iu-bremen.de}              
              }
              

   \abstract{In the framework of the Chandrasekhar-mass deflagration
model for Type Ia supernovae (SNe Ia), a persisting free parameter is
the initial morphology of the flame front, which is linked to the
ignition process in the progenitor white dwarf. Previous analytical
models indicate that the thermal runaway is driven by temperature
perturbations (``bubbles'') that develop in the white dwarf's
convective core. In order to probe the conditions at ignition
(diameters, temperatures and evolutionary timescales), we have
performed hydrodynamical 2D simulations of buoyant bubbles in white
dwarf interiors. Our results show that fragmentation occurring during the bubble rise affects the outcome of the bubble evolution. Possible implications for the ignition process of SNe Ia are discussed. 

 }

\maketitle


\section{Introduction}
\label{intro}

Advances in computer simulations of type Ia supernovae (SN Ia) have
led to a rough consensus over the nature of these explosions, even if
several issues remain controversial. SNe Ia are believed to be
the outcome of thermonuclear explosions of carbon-oxygen white dwarfs
(CO WDs) in binary systems.  Among possible explosion scenarios (see
\citealt{hn00} for a review of them) the single degenerate scenario
fares best in explaining the observed homogeneity of SNe Ia
\citep{l00}. In this model a CO WD approaches the Chandrasekhar mass
by mass transfer from a non-degenerate companion and is disrupted by a
thermonuclear explosion.

In this paper we wish to study the processes that lead to the ignition
of SNe Ia. The mechanism that ignites the thermonuclear flame in the WD
core is the link between the late stages of the progenitor evolution
and the development of the initial flame morphology. From a practical
point of view, it provides the initial conditions for the explosion
simulations. A large set of different initial conditions has been
tested in the literature of SN Ia simulations, with results ranging
from mildly energetic \citep{rhn02b,gko03,thr04,rh04} to failed explosions
\citep{cpv04} in pure deflagration explosion models. The initial flame
setup in multi-dimensional 
simulations of SN Ia explosions is considered basically a free
parameter, though constrained by some analytical works
(\citealt{gsw95,wwk04,ww04}; Sect.~\ref{bubbles}). Consequently
most simulation schemes are tested against different sets of initial
flame shapes, allowing a comparison of the results for various setups
\citep{rhn02b,rh04,gsb05,lah05}.

Among the possible initial conditions the most intuitive one, also for its
use in 1D simulations, is the centrally ignited flame
\citep[e.g.][]{nh95a,rhn99b,rhn02a,gko03}. A different ignition
condition is provided by a number of spherical flame kernels placed in
the innermost part of the WD and detached from the center.  Early
works in 2D \citep{nhw96,rhn99b} implemented only a few blobs per
quadrant, while newer simulations \citep{rhn02b,thr04} have a finer
spatial resolution and allow the implementation of more
seeds. Nevertheless, the maximum number of blobs that constitute the
initial flame is still constrained by resolution rather than
physics. This situation may be cured by the use of co-expanding
computational grids \citep{r04}. A hybrid approach between centrally
ignited flame and multi-spot scenario is presented by \citet{rh04} in
a full-star 3D simulation.

The influence of different initial conditions on the outcome of the
explosion has been discussed in several papers. There is general
agreement that an initially larger number of blobs produces more
vigorous explosions because the initial flame surface is relatively
larger and can provide more flame acceleration. Also the comparison of
different initial conditions in \citet{rh04} shows that the hybrid
``foamy'' initialization provides more seeds for the development of
instabilities and gives a larger total energy and production of
$^{56}\mbox{Ni}$ with respect to the model ignited at the center. In
the same work, the comparison with the failed explosion found by
\citet{cpv04} is addressed. Their simulation starts from a single
spherical flame seed, displaced slightly off-center.

Since the features of SN Ia explosions depend on the number of
igniting points in the WD core, an interesting question pertains to an
estimate of this number. In the papers cited above the study of this
problem was conducted by means of an \emph{a posteriori} evaluation of
the explosion features as a function of the initial conditions,
without exploring the physics of the ignition process itself (with the
exception of \citealt{wwk04}). A successful ignition theory has to
provide a physical basis for the choice of favored initial conditions in SNe
Ia explosion models.

In this paper, a new approach to the ignition problem is based on 2D
simulations performed with the FLASH code \citep{for00}. We will use
an indirect approach: the features of the ignition process will be
explored by studying temperature perturbations in the WD core
(``bubbles'') that are supposed to trigger the ignition
(cf.~Sect.~\ref{bubbles}). Using the results of the bubble
simulations, we are able to obtain important clues about the ignition
physics. 

The paper is structured as follows: in Sect.~\ref{ignition} the basic
ideas of the physics of ignition are introduced. Section \ref{numerical}
describes the setup of the simulations. Section \ref{simulations}
focuses on the bubbles, presenting the relevant physics and the issues
related to the 2D simulations. Some important parameters for the
bubble evolution are highlighted and their role is explored in
Sect.~\ref{parameter}. In Sect.~\ref{final-discussion} we use the
results of the simulations in order to study possible implications for
the ignition of SNe Ia. Our conclusions are summarized in
Sect.~\ref{conclusions}.


\section{The ignition process of SNe Ia}\label{ignition}

\subsection{Final stage of the progenitor evolution}\label{final}

According to stellar evolution models, CO WDs are the endpoints of the
evolution of main sequence stars in the mass range $M \approx 3 - 9
M_{\sun}$ \citep{uny99}. The energetics in the WD's interior during
the accretion phase is determined by the following processes:

\begin{itemize}

\item Compressional heating caused by accretion, described in detail
e.g.~by \citet{i82}. 

\item Nuclear burning: Compression and heating trigger hydrostatic
carbon burning in the WD core.  According to \citet{n82} the burning
starts in the pycnonuclear regime and turns to the thermonuclear
regime for $T > 5 \times 10^7\ \mathrm{K}$.

\item Neutrino emission: The energy loss by neutrino emission is
important at densities below about $2 \times 10^9\ \mathrm{g\
cm^{-3}}$ \citep{a71,nty84,ww86}. In these conditions, the dominant
contribution to neutrino losses comes from the production of plasmon
neutrinos, a mechanism described for example by \citet{c83} and
\citet{wsm04}. As a result of accretion, the density increases
above $2 \times 10^9\ \mathrm{g\ cm^{-3}}$, the thermal photons do not
have enough energy for plasmon excitation and the neutrino production
in the WD core is strongly suppressed.
\end{itemize}

In the evolutionary plane of central density and temperature
($\rho_\mathrm{c}\, ; T_\mathrm{c}$) (cf.~\citealt{yl03}), the locus
where the energy release due to carbon burning is equal to the energy
loss by neutrino emission is traditionally referred to as the ``ignition
line''. Actually it marks the start of the last phase of the
progenitor evolution. In order to get rid of the energy output of the
carbon burning, the WD develops a convective core whose evolution is
crucial for the knowledge of the ignition properties. The duration of
the convective phase is of the order of $10^3$ years
\citep{hn00,yl03}.

When the core temperature reaches $6-7 \times 10^8\ \mathrm{K}$, the
convective zone encompasses most of the WD mass.  Convection enters a
reactive regime \citep{wwk04}, in which the convective turnover
timescale and the nuclear timescale $\tau_\mathrm{n}$, defined as the
time required to significantly reduce the carbon abundance in an
isolated region of the fluid, are comparable. At $T \simeq 10^9\
\mbox{K}$ the critical temperature \citep{tw92} for carbon burning is
reached, i.e.~the temperature where the energy generation rate is equal
to the heat conduction rate.  This point defines the start of the SN Ia
explosion, the ignition of the thermonuclear runaway. 

Beyond the description of this sequence of events, a consistent
ignition model of SNe Ia is required to answer more detailed
questions.  In particular, the initial location and morphology of the
flame in the WD, the time evolution of the ignition process and the
thermodynamic state of the WD core prior to runaway need to be
explored. One possible approach to study the multi-dimensional
features of reactive convection is a direct numerical simulation of
this phase. A 2D simulation performed with an implicit code and
limited to the last few hours before the runaway was performed by
\citet{hs02}. According to the results of this work, the ignition
occurs at the center of the WD and is triggered by compressional
heating in the convective flow. These conclusions were challenged by
\citet{wwk04} who claim that the central ignition was either an
artifact caused by the forced symmetry of the 2D simulation, or caused
by insufficient resolution.


\subsection{Temperature fluctuations in the WD core}\label{bubbles}

The idea that the SN Ia explosion is ignited by floating bubbles has
been proposed first by \citet{gsw95}. The formation of these
temperature fluctuations in the WD's convective core is connected to
the turbulent behavior of the WD fluid and can be explained in terms
of velocity fluctuations. \citet{ww04} show that, when a fluid
element, during the convective motion, approaches the energy
generating core of the WD\footnote{Considering the high sensitivity of
the C-burning rate on temperature, it may be proved \citep{wwk04} that
one half of the energy is produced in a small (estimated mass of
$0.01\ M_{\sun}$, corresponding to about $130\ \mbox{km}$ for the WD
model examined by those authors) central part of the convective core.}
with a velocity smaller than the average convective velocity, it stays
there for a slightly longer time than average and becomes hotter (and
less dense) than the surrounding material. Therefore, the bubble is
subject to buoyancy and rises from the center of the WD.

The rising bubbles play a key role in the ignition because they are
regarded as the ``seeds'' for the subsequent flame propagation. The
initial flame in the progenitor is determined by motion of the bubbles
inside the convective core. The study of the bubble features is an
important part of investigating the ignition process.

\citet{gsw95} have developed an analytic model for the buoyant
evolution of burning bubbles. They have performed a parameter study in
which they investigate the bubble motion as a function of the initial
bubble position, the diameter, and the initial temperature excess. The
conclusion is that the bubbles reach the runaway temperature when they
are at a central distance in the range $100 - 250\ \mbox{km}$, moving
with a speed of about $100\ \mathrm{km\ s^{-1}}$.

The topic has been revisited by \citet{wwk04}. On the topic of
ignition they conclude that it is initiated by bubbles at a central
distance of up to about $150 - 200\ \mbox{km}$, when the central WD
temperature is about $7.7 - 7.9 \times 10^8\ \mbox{K}$. Moreover, they
claim that a multi-point ignition is possible. This results from the
fact that the e-folding timescale for the number of bubbles is
comparable with the time for the expansion in the SN Ia explosion to
shut off the ignition ($\sim 0.1\ \mathrm{s}$).

Finally, the ignition has been addressed in a more general, heuristic
model of turbulent convection by \citet{ww04}. Two convective flow
patterns are considered, an isotropic turbulent flow and a dipolar jet
flow. As far as the central WD temperature is concerned, the
conclusions for the SN ignition are similar to previous work. On the
location of the hottest points in the WD the authors consider the
competition between nuclear heating and adiabatic cooling and estimate
that the ignition points should be concentrated at about 100 km from
the WD center. For an isotropic flow it means that the ignition points
should occur in a \emph{shell} of 100 km of diameter while for a
dipole flow these points would be clustered rather along the flow
axes.


\section{Numerical tools}\label{numerical}

The 2D simulations which we will present here have been produced using
FLASH (version 2.3), a parallel adaptive-mesh hydro code
\citep{for00}. The FLASH code has been used, among other topics, for
the study of SN Ia explosions \citep{cpv04,pcl04} and, in a different context,
also in the study of the morphology of rising bubbles
\citep{rdr04,bhr05,brh05} in cooling flows in galaxy clusters. 

\begin{figure}
  \resizebox{\hsize}{!}{\includegraphics{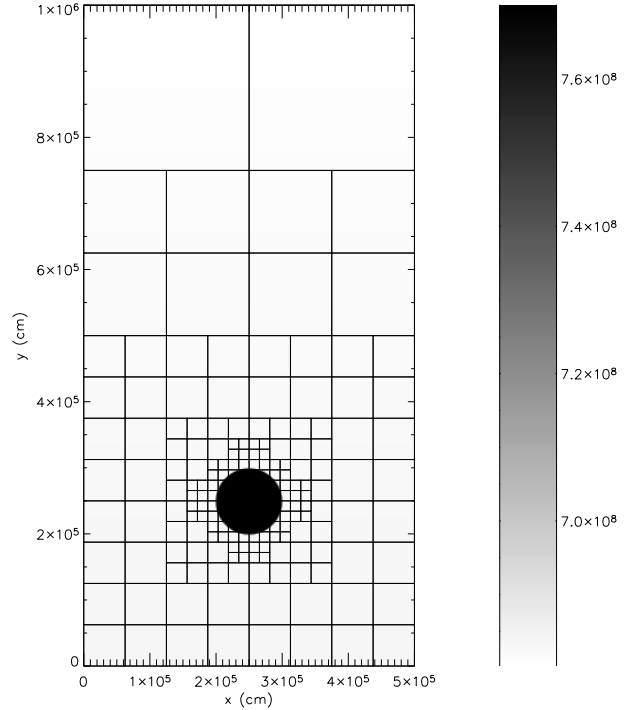}}
  \caption{Temperature plot, that shows a part of the initial setup of a bubble
simulation. The temperature is
coded in gray-scale, in Kelvin. The grid is
superimposed on the plot; each square is defined as a ``block'' of the
AMR structure and contains $8 \times 8$ computational cells. The plot
refers to a calculation with an initial bubble temperature of $7.7
\times 10^8\ \mbox{K}$, bubble diameter $1\ \mbox{km}$ and initial
distance from the center $100\ \mbox{km}$.}
  \label{domain-temp}
\end{figure}

Figure \ref{domain-temp} shows the computational domain at the
beginning of a 2D bubble simulation. As explained below, the
computational domain encloses only a small part of the WD. In this
domain the bubble is set at rest. Since most of the calculations are
performed with bubbles of $1\ \mbox{km}$ in diameter, the following
description will refer to this case (for different bubble diameters
one just has to scale the lengths accordingly).

The size of the computational domain is $5 \times 20\ \mbox{km}$. The
bubble is initialized as a temperature perturbation in pressure
equilibrium with the surrounding matter. The values of the
thermodynamical variables in the domain are obtained from a
one-dimensional WD model provided by S.~Woosley, with mass
$M_{\mathrm{WD}} \approx 1.38 M_{\sun}$, radius $R_{\mathrm{WD}}
\approx 1600\ \mbox{km}$, $T_{\mathrm{c}} = 7 \times 10^8\ \mbox{K}$,
$\rho_{\mathrm{c}} = 2.55 \times 10^9\ \mathrm{g\ cm^{-3}}$. The
extent of the convective zone in the model is about 1000 km.

Since the computational domain encloses a relatively small part of the
WD, a reasonable approach is to neglect the effect of curvature and to
map the data from the WD model in a plane-parallel approximation. The
values of physical quantities on the computational grid at coordinate
$y$ are taken from the quantities in the WD model at the radius $R -
y_{\mathrm{b}} + y$, where $R$ is the parameter which expresses the
initial distance of the bubble from the WD center and $y_{\mathrm{b}}
= 2.5\ \mbox{km}$ is the $y$ coordinate of the bubble center in the
computational domain. Before initialization, the pressure and density
data are slightly 
modified to ensure the hydrostatic equilibrium, as described by
\citet{zdz02}. A spatially constant gravitational force, computed from
the WD model, is applied to the computational domain. This assumption is not strictly valid because the variation of the gravitational acceleration along the whole computational domain (20\ \mbox{km}) is relevant. However, we do not expect this to affect our results since the extent of the bubble motion is always smaller than the computational domain, namely of the order of a few kilometers (Fig.~\ref{morphology}), comparable with the bubble diameter. The boundary
conditions are set to be reflecting in the upper and lower boundaries,
and periodic in the $x$-boundaries.

The simulations are performed in 2D Cartesian geometry. In principle,
this is not the most natural choice of geometry since a bubble is a
spherical object and it would be better represented in 2D in
cylindrical coordinates, exploiting the axial symmetry. Nevertheless,
some tests have been performed, both, in 2D Cartesian and
cylindrical geometry. Simulations in cylindrical geometry present
special difficulties, with respect to the Cartesian geometry, because
of larger numerical dispersion (cf.~Sect.~\ref{physics}) and the
appearance of ``axis jets'' which distort the bubble morphology.
The tests show that the choice of the Cartesian
geometry is acceptable for following the thermal evolution of bubbles.

The adaptive mesh refinement implemented in FLASH is apparent in
Fig.~\ref{domain-temp}. Resolution tests have shown that a good
compromise between computational cost and adequate resolution is to
use five levels of refinement yielding an effective grid size of $[256
\times 1024]$ zones (cf.~Sect.~\ref{physics}). This corresponds to a spatial resolution of $2
\times 10^3\ \mbox{cm}$ at the most resolved level.

Among the different equations of state implemented in FLASH, the
Helmholtz EOS described by \citet{ts00} was chosen for our
setup. In order to follow the hydrostatic carbon burning the small
reaction network {\tt iso7} was used. This $\alpha$-network is
adequate for the physical problem under examination, as confirmed by
\citet{thw00}.

To date, in the latest publicly available version of FLASH (v.~2.5)
there is no treatment of flame propagation. Therefore, the simulations
are followed until $T \approx 10^9\ \mbox{K}$, which is when the flame
is going to start, and are then stopped. Since this work is devoted to
the study of the progenitor's evolution \emph{before} the runaway
starts, this limitation does not concern us much.


\section{Two-dimensional simulations of rising
bubbles}\label{simulations}

\subsection{List of the performed simulations}\label{performed}

In order to cover the broad range of bubble parameters that is likely
to exist inside the WD, we present a study that explores the
dependence of the bubble evolution on the three main parameters:

\begin{itemize}
\item The initial bubble diameter, $D$.
\item The initial bubble temperature, $T$. 
\item The initial distance of the bubble from the WD center, $R$.
\end{itemize}

The aim of the study is to understand under which (physically
meaningful) conditions thermonuclear runaway ensues and what the related
timescales are.

\begin{table}
\caption{Scheme of the simulations which have been performed for the
parameter study.}
\centering
\begin{tabular}[t]{c c c c c } \hline \hline
Distance from the        &     &           & Parameters          &   \\
WD center and            &     &           & of the simulations  &   \\
background temperature   &     &           &                     &   \\ \hline
                         &     &           &                     &   \\
$ R = 50\ \mathrm{km}$   & 7.3 & {\bf 7.5} & {\bf 7.6}           &   \\
{\it Background} = 6.96  &     &           &                     &   \\
\hline 

                         &     &         &{\bf 7.7, $D$ = 5 km}& \\ \cline{4-4}
$ R = 100\ \mathrm{km}$  & 7.3 & 7.5     & 7.7       & {\bf 7.9} \\ \cline{4-4}
{\it Background} = 6.83  &     &         & 7.7, $D$ = 0.2 km     & \\ 
\hline
                         &     &           &                     &  \\
$ R = 100\ \mathrm{km}$  & 7.3 & 7.5       & 7.7                 & 7.9   \\
{\it Background} = 6.63  &     &           &                     &  \\ \hline 
\end{tabular}

\label{list}
\end{table}

The thirteen calculations performed for our parameter study are listed
in Table \ref{list}. In the first column the central distances of the
bubbles are given, together with the background temperature of the WD
at that distance from the center (in units of $10^8\ \mbox{K}$). In the
other four columns, the simulations are identified by the initial
temperature, again in units of $10^8\ \mbox{K}$. The bubble diameter
$D$ is fixed at 1 km unless explicitly specified. Simulations that
reach the thermal runaway are in boldface. To sum up, the initial
parameters have been varied in the following way:
\begin{enumerate}
\item $D$ (bubble diameter): for the case with $T = 7.7 \times
10^8\ \mbox{K}$, $R = 100\ \mbox{km}$, the three values 0.2, 1, 5 km
have been tested. The lower limit is constrained by
computational feasibility, decreasing the timestep of the simulations
together with the spatial resolution because of the CFL condition for
the equations of hydrodynamics.
\item $T$ (bubble temperature): $7.3 - 7.9 \times 10^8\ \mbox{K}$. The
upper end of this range is set in order to avoid too extreme
temperature contrasts with respect to the background. For bubble
temperatures smaller than the lower end of the range, the nuclear
timescale is longer than about 5 s. It seems unlikely that a bubble
can have such a long evolution, without being disrupted as described
in Sect.~\ref{physics}.  
\item $R$ (distance from the WD's center): the three values 50, 100, 150
km have been explored. Following \citet{ww04}, we assume that the
bubbles are produced in the energy generating core of the WD, which
extends approximately over this range of central distances.
\end{enumerate}


\subsection{Diagnostic quantities}\label{diagnostic}

First, we will define some quantities that are useful for the
interpretation of our simulations.

A first difficulty in defining these quantities is the lack of a good
criterion about the bubble location in the computational domain. This
is of course a typical problem, when dealing with an Eulerian system
of coordinates. In principle, hot material (the bubble is set to be
rather hotter than the background) could keep track of the bubble
evolution. Unfortunately, in Sect.~\ref{physics} we show that
numerical diffusion hampers this approach and one cannot make safe use
of ``bubble averaged quantities''. Hence, the description of the
thermal evolution of the bubble will be done by studying the maximum
temperature in the bubble.

A useful quantity for the following analysis of the bubble is its
area. As written above, it is difficult to 
provide an unambiguous definition of this variable. Test simulations
have established that a good criterion is given by the following expression: 

\begin{equation}
\label{bubble-area}
\mathcal{A}_{\mathrm{bubble}} = \sum_{bubble} \mathcal{A}_{\mathrm{zone}}
\end{equation}
where the sum is calculated over the zones where $T_{\mathrm{zone}} >
T_{\mathrm{threshold}}$ and the threshold temperature is defined by

\begin{equation}
T_{\mathrm{threshold}} = \left\{ \begin{array}{rcl} 0.95 \cdot
T_{\mathrm{bubble,ini}} & \mbox{for} & T_{\mathrm{bubble,ini}} >  7.5 \times
10^8\ \mathrm{K}) \\
 0.96 \cdot
T_{\mathrm{bubble,ini}} & \mbox{for} & T_{\mathrm{bubble,ini}}
\leqslant  7.5 \times 
10^8\ \mathrm{K}) \end{array} \right.
\end{equation}
where $T_{\mathrm{bubble,ini}}$ is the initial bubble temperature. The
quantity $\mathcal{A}_{\mathrm{zone}}$ is geometry dependent. In Cartesian
geometry it can be simply identified with the zone area
$\mathcal{A}_{\mathrm{zone}} = dx \cdot dy$, where $dx$ and $dy$ are the zone
sizes in the two directions. 

It is also useful to introduce the effective gravitational acceleration
\citep{gsw95,wwk04}. The acceleration $g_{\mathrm{eff}}$ exerted by
buoyancy on a  
rising bubble is proportional to its density contrast:

\begin{equation}
\label{g-effective}
g_{\mathrm{eff}}(r) = g(r)\, \frac{\Delta \rho}{\rho} = \frac{G\,
M(r)}{r^2}\, \delta_{\mathrm{p}}\, \frac{\Delta T}{T} \; .
\end{equation}

In the previous equation, the density contrast between the bubble and
the background $\Delta \rho / \rho$ can be expressed in terms of the
temperature contrast $\Delta T / T$ times $\delta_{\mathrm{p}}$, the
logarithmic derivative of density with respect to temperature at
constant pressure.


\subsection{The physics of the bubble}\label{physics}

There exists an extensive literature on the fluid mechanics of rising
bubbles. Nonetheless, we are not aware of any work that addresses the
peculiarities of the present setup (degenerate matter, very small
density contrasts and nuclear burning).
 
\begin{figure}
  \resizebox{\hsize}{!}{\includegraphics{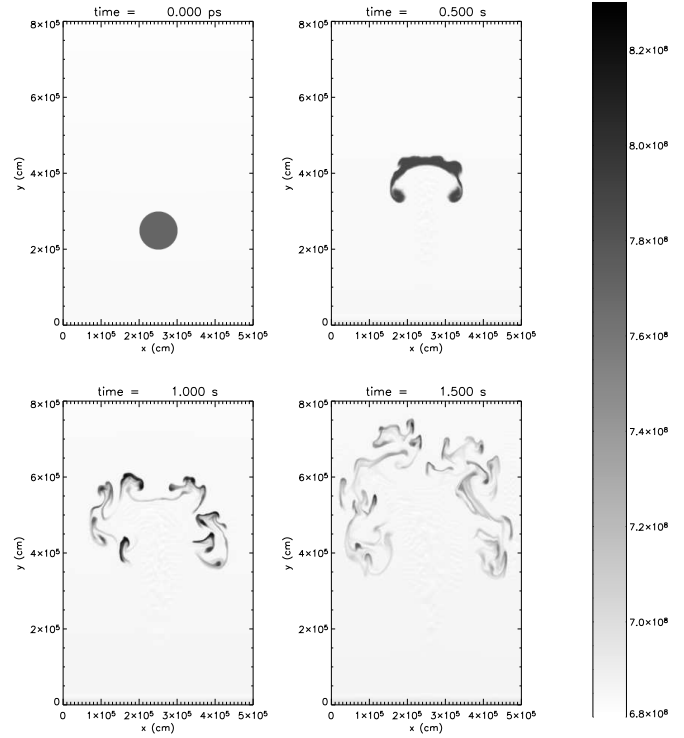}}
  \caption{Series of temperature plots, showing the evolution of a
bubble in a simulation with the initial parameters $T = 7.7 \times 10^8\
\mbox{K}$,  $D = 1\ \mbox{km}$, $R = 100\ 
\mbox{km}$. The temperature is coded in gray-scale, in Kelvin.}
  \label{morphology}
\end{figure}

At first, our discussion will focus on the following ``reference
choice'' of parameters: initial temperature $T = 7.7 \times 10^8\
\mbox{K}$, initial diameter $D = 1\ \mbox{km}$, initial central
distance $R = 100\ \mbox{km}$ (Fig.~\ref{morphology}). These values have
been chosen to lie well inside the range of variation of the three parameters,
discussed in Sect.~\ref{performed} and also suggested by \citet{wwk04}. 
First we will describe the results for these reference parameters, while in
Sect.~\ref{parameter} we will explore the results of varying these
parameters.

If we neglect adiabatic cooling (as motivated in Sect.~\ref{remarks}),
the evolution of the bubble depends only on the interplay between nuclear
burning and hydrodynamical instabilities.

Hydrostatic carbon burning sets the timescale for the bubble to reach
the thermonuclear runaway. This nuclear timescale $\tau_{\mathrm{n}}$
\citep{wwk04} has been introduced in Sect.~\ref{final}.  Using the
evolutionary timescales of the simulations that reach a runaway
(cf.~Table~\ref{list}) and adding data from further tests an
analytical fit to $\tau_{\mathrm{n}}$ has been found. The result is

\begin{equation}
\label{my-fit}
\tau_{\mathrm{n}} \approx 10\ \left( \frac{7}{T_8}
\right)^{22}\ \left( \frac{2.5}{\rho_9} \right)^{4}\ \mbox{s}
\end{equation}
where $T_8 = T / (10^8\ \mathrm{K})$ and $\rho_9 = \rho / (10^9\
\mathrm{g\ cm^{-3}})$. This fit of $\tau_{\mathrm{n}}$ is in rough agreement
with the derivation by \citet{wwk04} for the conditions of interest in
the WD core.

The bubble motion is a special case of a system subject to the
Rayleigh-Taylor instability (in the following, RTI). In a
gravitational field with the vector of gravitational acceleration
$\vec{g}$ pointing downwards along the $y$-axis, the hotter and
lighter fluid accelerates upwards. The development of the RTI and the
features of the motion in this situation have been addressed in
several works \citep{dt50,t50,l55,gl88,l96}. A crucial parameter in
this problem is the Atwood number $At$,

\begin{equation}
At = \frac{\rho_2 - \rho_1}{\rho_2 + \rho_1}
\end{equation}
where $\rho_1$ and $\rho_2$ are the bubble and background densities,
respectively. In our bubble problem, the density contrast is very small 
because of the degeneracy of the WD matter, and $At$ is of
the order of $10^{-4}$. Theory predicts that the rise velocity
of the bubble tends to a value $v_{\mathrm{b}}$ that is
determined by the equilibrium between buoyancy and drag forces. An
analytical model developed for nonlinear, single-mode, classical RTI
at arbitrary Atwood number by \citet{g02} provides the estimate

\begin{equation}
\label{goncharov}
v_{\mathrm{b}} = \sqrt{\frac{2At}{1+At}\
\frac{g}{Ck}}\; ,
\end{equation}
where $k \sim 2 \pi / (D/2)$ is the wavenumber of a perturbation of
size of order of the bubble radius $D/2$, and $C$ is a numerical
constant whose value is 3 in 2D and 1 in 3D.  From
Eq.~(\ref{goncharov}) one gets, for the reference choice of bubble
parameters, $v_{\mathrm{b}} \simeq 1.3 \times 10^5\ \mathrm{cm\
s^{-1}}$. This is an appropriate order-of-magnitude estimate for the
bubble velocity in the simulations, even if the morphology evolution
of the bubble prevents a more quantitative analysis.

The vortical motions produced during the bubble rise cause its fragmentation as it is clear from
Fig.~\ref{morphology}. In the nonlinear stage of the RTI a large variety of phenomena can account for the observed fragmentation, as described e.g.~by \citet{agl03} and reviewed by \citet{nt96} in the framework of inertial confinement fusion. We will not address a detailed study about the nature of these secondary instabilities. The features of the fragmentation process and some related numerical issues are described below.

An approximate estimate for the timescale of 
the fragmentation process can be given by defining a dispersion timescale

\begin{equation}
\label{tau-rti}
\tau_{\mathrm{disp}} = \frac{D}{v_{\mathrm{b}}}
\end{equation}
which, with the reference values of the parameters and the use of
eq.~(\ref{goncharov}), gives a timescale of about $0.8\ \mbox{s}$, a
sort of ``bubble lifetime'', to be compared with the morphological
behavior shown in Fig.~\ref{morphology}. 

\begin{figure}
  \resizebox{\hsize}{!}{\includegraphics{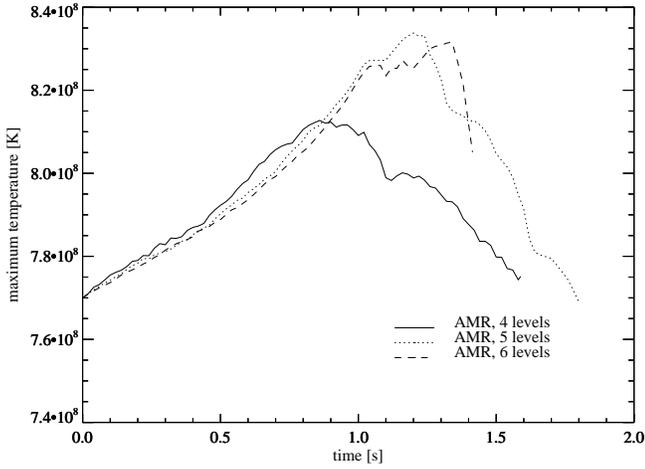}}
  \caption{Evolution of the maximum bubble temperature in simulations
performed with the reference choice of parameters and increasing resolution, indicated by the different levels of refinement.}
  \label{maxtemp}
\end{figure}

\begin{figure}
  \resizebox{\hsize}{!}{\includegraphics{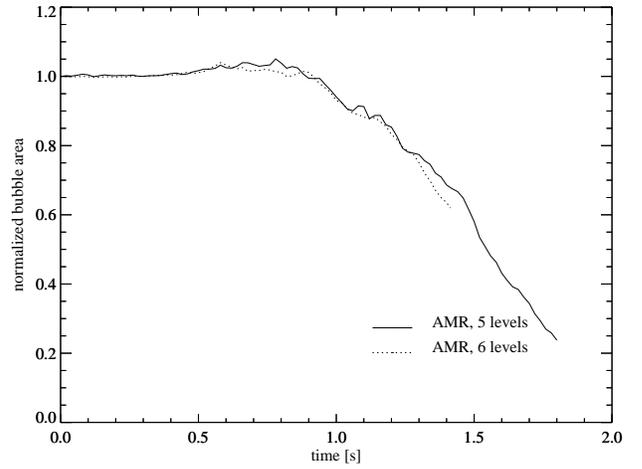}}
  \caption{Evolution of the normalized bubble area for the reference
choice of parameters, in a comparison among simulations with different
resolution.} 
  \label{area}
\end{figure}

During the fragmentation, the
typical length scale of the bubble fragments decreases in time. From
a physical point of view, one can expect that the dispersion goes on
until the typical length scale of a bubble part is comparable with the
minimum size $\lambda_{\mathrm{min}}$, defined as the diameter of the
bubble in which the heat generated by nuclear burning is balanced by
heat diffusion. An evaluation of this quantity is performed, with
different derivations, by \citet{wwk04} and \citet{gsb05}, and gives
results in the range $10 - 100\ \mbox{cm}$.

For bubbles whose size is larger than $\lambda_{\mathrm{min}}$, the
nuclear heating is larger than the heat loss by conduction. In these
bubbles the temperature increases (in this phase, in principle, they
can still reach $10^9\ \mbox{K}$ and trigger the thermonuclear
runaway), until disruption causes their typical size to decrease to
$\lambda_{\mathrm{min}}$. Then they begin to cool down. Thus, the
trend of the maximum temperature shown in Fig.~\ref{maxtemp} can be
easily interpreted: the bubble is heated until either runaway
occurs or the bubble is disrupted.

\begin{figure}
  \resizebox{\hsize}{!}{\includegraphics{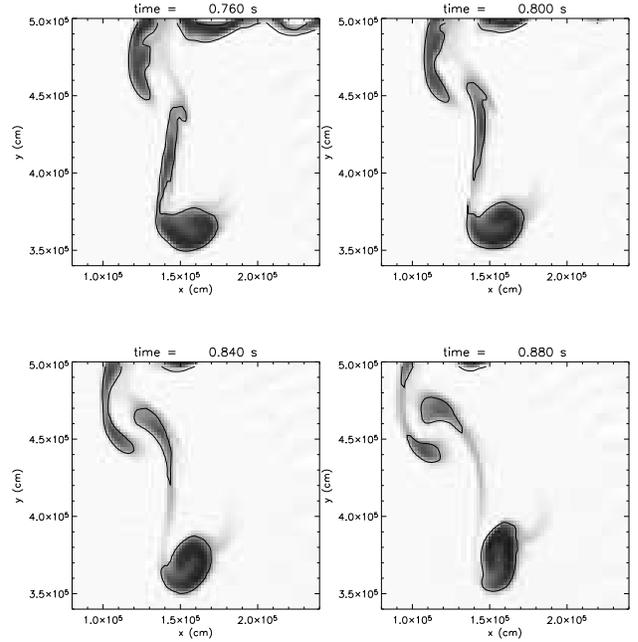}}
  \caption{Series of temperature plots referring to the same
simulation as Fig.~\ref{morphology}, showing the evolution of a
bubble detail. The contours in the plots
mark the bubble area, according to the definition
of Eq.~\ref{bubble-area}.}
  \label{detail-dispersion}
\end{figure}

One should note, however, that the length scale $\lambda_{\mathrm{min}}$
for dispersion driven by heat conduction is much smaller than the
spatial resolution of the simulations ($2 \times 10^3\
\mbox{cm}$). $\lambda_{\mathrm{min}}$ is so small that a direct
simulation to that scale, even with AMR, is not feasible. What is
actually observed in the bubble simulations is not a physical but a
\emph{numerical} dispersion. Figure \ref{detail-dispersion} is meant to
clarify this important point. The plots are zoomed on a small detail,
a sort of ``bridge'' linking the top of the bubble with a lower
vortical structure. This detail becomes thinner and, as its width is
approximately 2--3 times the spatial resolution, it cools off
unphysically. It is thus lost from the bubble area according to the
definition in Sect.~\ref{diagnostic}.

The dependence of the bubble fragmentation process on resolution is shown in Fig.~\ref{maxtemp}. As expected, the numerical dispersion depends on spatial resolution, and the temperature decrease occurs earlier for the simulation with coarser level of refinement. The resolution which has been chosen for the parameter study (five levels of refinement) guarantees however to follow reliably the fragmentation process, as shown in Fig.\ref{area}. Interestingly, the area evolution in the two most resolved simulations does not show any significant dependence of the bubble area decrease on the level of
refinement of the simulation. Probably, this happens because
the hydrodynamical instabilities in the more resolved calculation
produce structures that are globally more numerous, though smaller in
size.

As described above, electron conduction is an important ingredient in the bubble physics, but can be neglected at the length scales which are explored in this study. It is known from the theory of thermonuclear combustion \citep[see e.g.~][]{tw92} that heat exchange can inhibit the growth of instabilities which are smaller than a minimum length scale $\lambda_{\mathrm{cond}}$. According to \citet{tw92}, 

\begin{equation} 
\label{lambdacond} 
\lambda_{\mathrm{cond}} = 4 \pi v^2 \left(g \frac{\Delta \rho}{\rho}\right)^{-1}
\end{equation} 
where $g$ is the modulus of the gravitational acceleration and $\Delta \rho / \rho$ is the density contrast of the bubble with respect to the background. The flame velocity $v$ can be meaningfully replaced, in the case of volume burning, by $\lambda_{\mathrm{min}} / \tau_{\mathrm{n}}$.  
In the case under examination ($T_8 = 7$, $\rho_9 = 2.5$) the typical values of the quantities in Eq.~(\ref{lambdacond}) are (cfr.~Sect.~\ref{physics}) $\tau_{\mathrm{n}} = 10\ \mathrm{s}$, $\lambda_{\mathrm{min}} = 100\ \mathrm{cm}$, $\Delta \rho / \rho \simeq 10^{-4}$ and $g \simeq 10^{10}\ \mathrm{cm\ s^{-2}}$. 
This estimate shows that $\lambda_{\mathrm{cond}}$ is of the order of $10^{-3}\ \mathrm{cm}$. It is smaller than $\lambda_{\mathrm{min}}$, therefore the described bubble physics is not affected by the stabilization. 


\subsection{Outcome of the bubble evolution}\label{motion} 

The following overview sums up the main features of the bubble
evolution:

\begin{itemize}
\item The initial bubble temperature determines the nuclear timescale
$\tau_\mathrm{n}$ (Eq.~\ref{my-fit}). This is the upper limit for the
duration of the bubble evolution.

\item The bubble is fragmented during its motion on a
timescale approximately given by $\tau_\mathrm{disp}$
(Eq.~\ref{tau-rti}). Physically, the dispersion proceeds down to
length scales where heat conduction is effective in dissipating the
energy generated by nuclear burning.

\item The previous argument is made more complicated by numerical
diffusion. It is not possible to quantify exactly the whole duration
of the physical dispersion phase described previously because this
process can be followed in the simulations only until the typical
length scale of the bubble substructures are comparable to 2--3 times
the spatial resolution.
\end{itemize}

The competition between nuclear heating and dispersion is the key to
understand the outcome of the bubble evolution and, consequently, also
to the ignition process of SNe Ia. Roughly speaking, if
$\tau_\mathrm{n}$ is smaller than $\tau_\mathrm{disp}$ the burning
prevails and the bubble reaches the thermonuclear runaway. Conversely,
if $\tau_\mathrm{n} \ge \tau_\mathrm{disp}$ the bubble is disrupted and
cools down. Analytical studies on ignition that is driven by floating
bubbles have not explored the bubble dispersion
\citep{gsw95,w01,wwk04,ww04}.

\subsection{Background turbulence}
\label{turb}

In all our simulations the background state of the WD was assumed to
be ``quiet'' and in hydrostatic
equilibrium. However, the convective flow in the WD 
prior to runaway is turbulent, with an estimated Reynolds number about
$10^{14}$ \citep{wwk04}. The integral length scale $L$, at which the turbulent
kinetic energy is injected, is given by the pressure scale height,
about $450\ \mathrm{km}$ for the WD model used here \citep{w01}. Assuming
Kolmogorov scaling for the typical velocity of the convective
eddie of length scale $l$, the characteristic turbulent velocity $U$
at this length scale is expressed by $U(l) \approx U(L)
(l/L)^{1/3}$. 

In our case, $l = 2 \times 10^3\
\mathrm{cm}$ (spatial resolution of the simulation at the most
resolved level) and $U(L)$ is in
the range $50 - 100\ \mathrm{km\ s^{-1}}$, from estimates based on
mixing length theory. Therefore, the typical turbulent velocity $U(l)$
is in the range $1.7 - 3.5\ \mathrm{km\ s^{-1}}$, which is of the same order of
magnitude as the typical rise velocity of the bubble, according to
Eq.~(\ref{goncharov}). A test simulation, performed by imposing a random
velocity field with an amplitude of the order of $U(l)$, shows that this
stirring affects the bubble evolution only marginally, contributing to
a moderately enhanced bubble disruption.
We conclude that this turbulence-induced bubble disruption can be considered
a second-order effect 
in the description of the relevant bubble physics discussed in
Sect.~\ref{physics}.

Our simulations are affected by
numerical dispersion and hence overestimate the
bubble disruption and the subsequent bubble cooling. From a physical
point of view one expects that the 
bubble temperature would increase for a longer time than is shown, for
example, in Fig.~\ref{maxtemp}. However, the numerical dispersion may
not be too unphysical because it may mimic the effect of the
turbulent-induced bubble disruption. A qualitative comparison of 
these two effects cannot be performed with our computational tools.


\section{The parameter study}\label{parameter}

\subsection{The effect of the bubble diameter $D$}\label{parameter-d}

\begin{figure}
  \resizebox{\hsize}{!}{\includegraphics{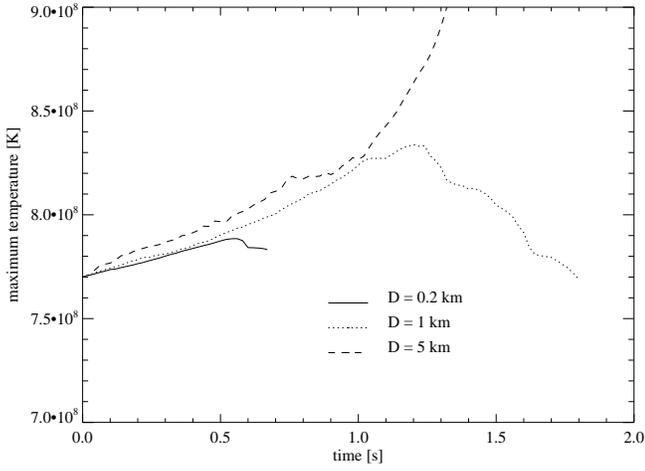}}
  \caption{Comparison of the evolution of maximum temperature in
simulations with $R = 100\ \mbox{km}$, $T = 7.7 \times 10^8\
\mbox{K}$ and $D$ indicated in the legends.}
  \label{diameter-study-maxtemp}
\end{figure}

\begin{figure}
  \resizebox{\hsize}{!}{\includegraphics{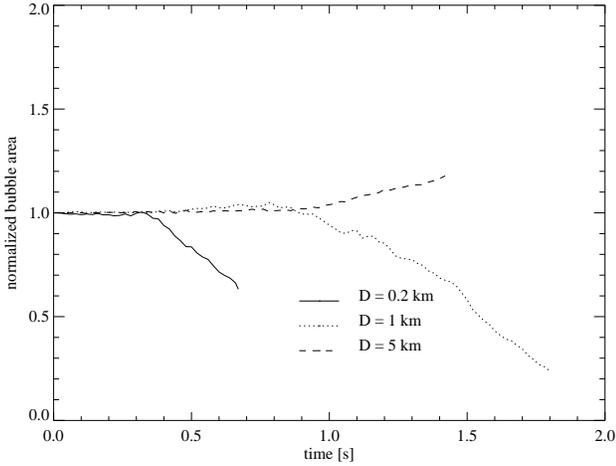}}
  \caption{Same as Fig.~\ref{diameter-study-maxtemp}, but the
evolution of the normalized bubble area is shown.}
  \label{diameter-study-area}
\end{figure}

As outlined in Table \ref{list}, the role of different bubble
diameters was studied on the basis of three calculations, with initial
parameters $T = 7.7 \times 10^8\ \mbox{K}$ and $R = 100\
\mbox{km}$. The initial diameter $D$ was varied by a factor of five
above and below the reference value, i.e.~$D = 0.2$, $1$ and $5\
\mbox{km}$. The extent of the computational domain and the spatial
resolution were scaled accordingly in order to resolve the initial
bubble by an identical number of zones. Figures
\ref{diameter-study-maxtemp} and \ref{diameter-study-area} present a
comparison of the maximum temperature and the normalized bubble area,
respectively.

As expected, the bubble diameter affects the dispersion
timescales. Indeed, the dispersion timescale (Eq.~\ref{tau-rti}) scales as
$D\, ^{1/2}$ since, from Eq.~(\ref{goncharov}), $v_{\mathrm{b}}
\propto D\, ^{1/2}$ as well. This scaling is nicely confirmed by the
comparison of the area evolution in
Fig.~\ref{diameter-study-area}. While in the simulation with
$D = 1\ \mbox{km}$ the area starts to decrease at approximately $t \approx 0.8\
\mbox{s}$, one can see that for the bubble with $D = 0.2\ \mbox{km}$
the analogous decrease starts at $t \approx 0.8 \cdot (0.2 /
1)^{1/2} \simeq 0.35\ \mbox{s}$.

For the same reason, the bubble with $D = 5\ \mbox{km}$ has a larger
$\tau_{\mathrm{disp}}$. Since the nuclear timescale does not depend on
the bubble diameter, this leads to $\tau_{\mathrm{disp}} >
\tau_{\mathrm{n}}$, and the bubbles goes to runaway.

From this study, one can conclude that the larger bubbles are more
likely to go to runaway. However, \citet{wwk04} find that turbulent dispersion limits the bubble size in the WD to $D \sim 1\ \mbox{km}$.


\subsection{The effect of the bubble temperature
$T$}\label{parameter-t}

\begin{figure}
  \resizebox{\hsize}{!}{\includegraphics{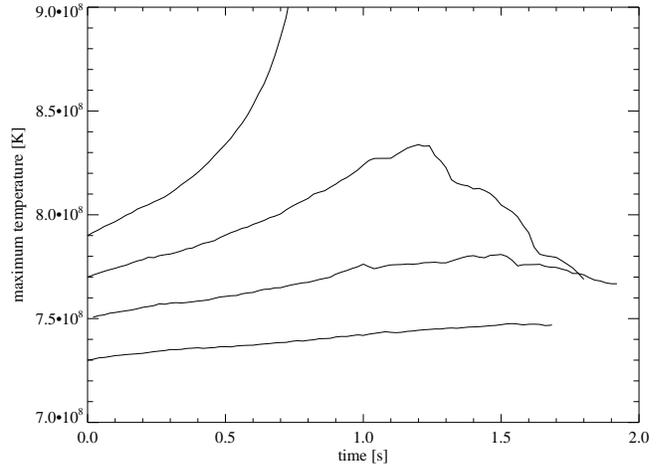}}
  \caption{Comparison of temperature evolution in
simulations with $R = 100\ \mbox{km}$, $D = 1\ \mbox{km}$ for four
  different initial temperatures.}
  \label{temp-study-maxtemp}
\end{figure}

\begin{figure}
  \resizebox{\hsize}{!}{\includegraphics{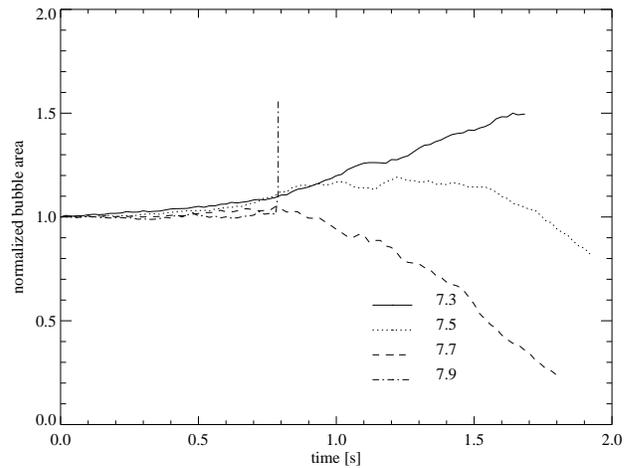}}
  \caption{Comparison
of the normalized bubble area in the parameter study on
temperature. The initial bubble temperatures are 
indicated in the legend, in units of $10^8\ \mbox{K}$.}
  \label{temp-study-area}
\end{figure}

The initial temperature $T$ is the most interesting parameter because
its role in the bubble physics is manifold. As shown in table
\ref{list}, several values have been tested. For sake of simplicity, a
discussion will only be presented for the simulations with $R = 100\
\mbox{km}$, $D = 1\ \mbox{km}$ and $T$ ranging from $7.3$ to $7.9
\times 10^8\ \mbox{K}$, keeping in mind that the inferred trends are
valid also for calculations with other central distances. Figures
\ref{temp-study-maxtemp} and \ref{temp-study-area} present a
comparative analysis of the evolution of maximum temperature and area.

The nuclear timescale depends on the bubble temperature in a rather
steep way (Eq.~\ref{my-fit}). Also the dispersion timescale
$\tau_{\mathrm{disp}}$ depends implicitly on $T$, approximately via the
square root of the Atwood
number $At$ (cf.~Eq.~\ref{goncharov}). $At$ varies
in the explored temperature 
range from $2.2 \times 10^{-4}$ ($T = 7.3 \times 10^8\ \mbox{K}$) to
$5.3 \times 10^{-4}$ ($T = 7.9 \times 10^8\ \mbox{K}$). In
simpler terms, the temperature contrast between the bubble and the
surrounding material is linked to the density contrast and hence to
the effective gravitational acceleration (Eq.~\ref{g-effective}). With
equal background 
temperature (i.e.~equal central distance $R$) hotter bubbles
experience larger accelerations and thus faster dispersion. On the
other hand, the nuclear timescale decreases even faster with
temperature. This indicates the existence of a threshold temperature
above which a bubble goes into runaway. At $R = 100\
\mbox{km}$ the threshold is found to be approximately $T = 7.9 \times
10^8\ \mbox{K}$.     


\subsection{The effect of the central distance $R$}\label{parameter-r}

\begin{figure}
  \resizebox{\hsize}{!}{\includegraphics{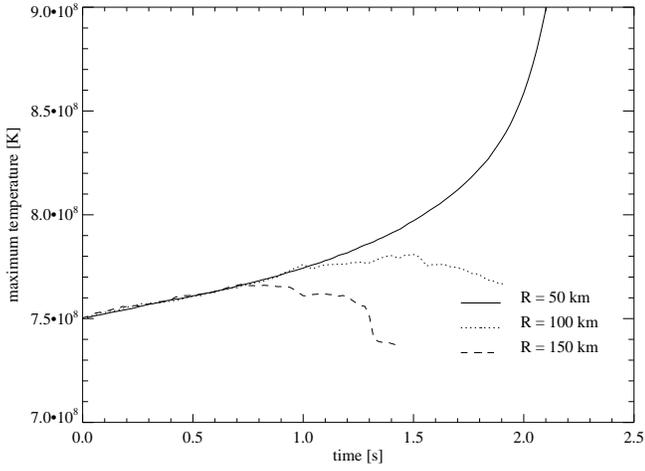}}
  \caption{Comparison of the evolution of  maximum temperature in
simulations with $T = 7.5 \times 10^8\
\mbox{K}$, $D = 1\ \mbox{km}$ and $R$ indicated in the legends.}
  \label{r-study-maxtemp}
\end{figure}

\begin{figure}
  \resizebox{\hsize}{!}{\includegraphics{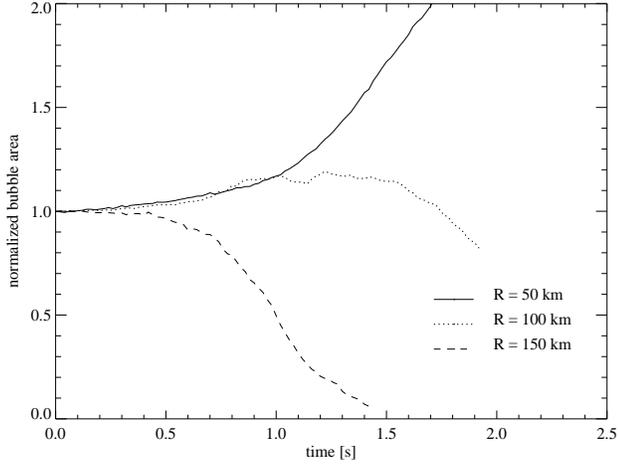}}
  \caption{Same as Fig.~\ref{r-study-maxtemp}, but the evolution of
the normalized bubble area is shown.}
  \label{r-study-area}
\end{figure}

While the parameter $T$ was shown to be the most interesting one for
the bubble physics, the central distance $R$ is probably the most
relevant one for the ignition scenario of SNe Ia. The analysis was performed by
comparing three simulations with the initial parameters $T = 7.5 \times 10^8\
\mbox{K}$, $D = 1\ \mbox{km}$ and $R$ equal to 50, 100 and
$150\ \mbox{km}$.

The increase of $R$ in the calculations affects the effective
gravitational acceleration for two reasons. First, the modulus of the
acceleration $g(R)$ increases with $R$ in the considered range of
distances. Second, the temperature
profile of the WD is decreasing (the background temperatures at
various $R$ are given in Table \ref{list}). So at equal bubble
temperature, the temperature (and density) contrast is increasing
with $R$. This implies that the effective gravitational
acceleration increases outwards, too. 

This effect is evident in the comparison between the timescales of the
two extreme cases, $R = 50$ and $150\ \mbox{km}$, from the analysis of
Figs.~\ref{r-study-maxtemp} and \ref{r-study-area}. The simulation
with $R = 50\ \mbox{km}$ goes to thermonuclear runaway. Its area
evolution shows an expansion (partly due to a problem of the
definition of ``bubble area'' when the temperature contrast between
the bubble and the background is relatively small). On the other hand,
as predicted the dispersion timescales of the other simulations are
noticeably shorter as $R$ increases.

The role of an increasing $g_{\mathrm{eff}}$ is also relevant for the
value of the threshold temperature at which the bubbles at different
initial distances from the center go into runaway. As shown in table
\ref{list}, the bubble at $T = 7.5 \times 10^8\ \mbox{K}$ reaches the
ignition temperature during the simulation with $R = 50 \ \mbox{km}$,
but with $R = 150 \ \mbox{km}$ even the bubble at $T = 7.9 \times
10^8\ \mbox{K}$ is dispersed before reaching the runaway.


\section{Discussion}\label{final-discussion}

\subsection{Further remarks on the physics of rising
burning bubbles}\label{remarks}

One of the most striking differences between the simulations presented
here and the first analytic study on bubble evolution by \citet{gsw95}
is that the latter does not take into account the dispersion of the
bubbles. Consequently, in their work the bubble velocity and the path
covered during the rise are much larger than in the present
study. Moreover, without dispersion the rise times are rather long, up
to about 25 seconds.

The rise velocity of the bubbles in the simulations is of the order of
$1\ \mathrm{km\ s^{-1}}$ (cf.~Eq.~\ref{goncharov}). Because the
timescale until the bubbles either go into runaway or are disrupted
(1--2 s) is so short, the bubble evolution occurs mostly near their
point of origin (Fig.~\ref{morphology}). However, the previous
argument neglects the convective motions in the WD core.

Placing a bubble at some distance $R$ from the WD's center and letting
it rise is a gross simplification because the convective velocity is
always much larger (in the range $50 - 100\ \mathrm{km\ s^{-1}}$, from
estimates based on the mixing length theory) than the simulated
buoyant velocity. So it is the convection which determines the motion
of the bubbles, rather than the rise velocity. Clearly, the convective
flow pattern crucially affects the spatial distribution of bubbles at
runaway.

As the bubble is advected by the convection, it may travel over a
distance that is long compared to the estimated central distance at
ignition, undergoing some cooling \citep{ww04}. The bubble
simulations cannot directly address this physical process. In our
setup, the bubble is initially at rest and no convective velocities
are imposed on the background. As a result the bubble travels only a
distance that is small compared to its diameter. One can prove that
adiabatic cooling is negligible in this case, as confirmed by test
simulations in which the nuclear burning was switched off.


\subsection{Implications for the theory of ignition of SNe
Ia}\label{implications}

In this section, our results are applied to the ignition process in SNe Ia.

First, a rough estimate of the extent of the size of the ignition zone
can be made on the basis of bubble lifetimes and convective
velocities. Taking the values of $1\ \mbox{s}$ and $70\ \mathrm{km\
s^{-1}}$, respectively, one finds that the extent of the motion of the
bubbles advected in the convective flow is about $70\
\mbox{km}$. Considering that the size of the 
energy-producing core is of the order of $100\ \mbox{km}$, an
estimate for the distance covered by a rising bubble before it triggers
the ignition is of the order of $150\ \mathrm{km}$. The uncertainties in the
convective velocity (in the range $50 - 100\ \mathrm{km\ s^{-1}}$) and in the
initial bubble location inside the WD core lead to some dispersion
around this value, which is of the order of $100\ \mathrm{km}$ (a
fraction of the pressure scale height). We note that 
this estimate is consistent with \citet{gsw95}, \citet{wwk04} and
\citet{ww04}.

It is clear that our approach for the study of ignition conditions
does not allow any conclusions about the departures from central
symmetry. This issue has to be studied with
other numerical tools. A promising way seems to be the use of 3D
anelastic simulations, adopted by \citet{kwg03} in the study of
convective flows in massive stars. Interesting results of this
technique, applied to SN Ia progenitors, are presented by
\citet{kwg05}. In any case, the present work highlights the importance
of the progenitor evolution for the initial conditions of the SN Ia
explosion. The convective pattern plays a crucial role in the ignition
process. The initial flame location depends mostly on the position
of the bubbles, which, in turn, are advected by the convective
motions inside the WD.

Another interesting issue is the initial number of igniting
points. Though our approach
for the study of the ignition process is indirect, some
useful hints can be obtained.  

In the present parameter study, several values of the initial
temperature have been tested without making assumptions concerning the 
probability distribution function (PDF) of the temperature
fluctuations. \citet{wwk04} investigate this problem and indicate two 
possible PDFs, depending on details on the convective mixing in the WD
core. In their work the PDF is either exponential or
Gaussian. Without making a choice among these two models, from the
generation process of the temperature perturbations it is intuitive
that bubbles with a relatively large temperature contrast with respect to
the background are less likely to be generated. 

Here a thought experiment of two different ignition scenarios is
useful. In the first case, let a hot bubble ($T \gtrsim 
7.5 \times 10^8\ \mbox{K}$) be in the core of a WD with central
temperature $T_{\mathrm{c}} \simeq 7.0 \times 10^8\ \mbox{K}$. The
estimated nuclear timescale is about $2\ \mbox{s}$. If this bubble is
on the ``hot tail'' of the temperature PDF, one can assume that the
probability of generating other bubbles with this $T$ (or hotter) within
$\tau_{\mathrm{n}}$ is not large, while colder bubbles would not have
time to reach the runaway temperature. In this scenario, the ignition
would occur in one (or a few) igniting points. However, considering
that the dispersion in more effective for bubbles of a large
temperature contrast with the background, such kind of ignition
scenario is likely to fail. If the explosion is not initiated, the WD
goes on increasing its central temperature.

The other scenario is directly connected with the previous idea. In
this second idealized setup, let the WD have a larger central
temperature, $T_{\mathrm{c}} \gtrsim 7.5 \times 10^8\ \mbox{K}$, and
the temperature fluctuations be relatively mild with respect to the
background. Under these assumptions, the estimated $\tau_{\mathrm{n}}$
of the bubbles is about $1\ \mbox{s}$, a value compatible with
multi-spot ignition, since the e-folding time of the number of bubbles
in the WD core is about 0.1 s \citep{wwk04}. Moreover,
these perturbations are not supposed to be very hot compared to
the background. The probability for such bubbles to be generated,
according to whatever PDF, should be large, allowing the presence of
very many of them in the WD core.

From these arguments the multi-point scenario emerges as a good
candidate for an ignition model of SNe Ia.


\section{Conclusions}\label{conclusions}

We have studied the ignition process in SNe Ia by simulating the
evolution of buoyant bubbles in the WD core. After discussing the
bubble physics, the role of the relevant parameters was studied by
means of a grid of thirteen simulations. This parameter study shows that floating bubbles may either ignite almost in place or be dispersed by fragmentation.
The previous results found an application in the study of ignition process in SNe Ia. The discussion in Sect.~\ref{implications} provides some clue that the multi-spot ignition model is the favored one. This 
bubble distribution is consistent with the initial conditions assumed
in many recent 3D simulations of SNe Ia explosions.

As pointed out e.g.~by \citet{hn00}, a successful model for
SNe Ia should explain, both, the homogeneity of the class and,
hopefully as a function of one or few parameters, the diversity of
observational features. The present study indicates that the
ignition process contributes to the homogeneity of SNe Ia.
The explanation comes from examining the second scenario described in
Sect.~\ref{implications}. According to the PDFs (no matter which one),
there is some probability for the occurrence of one (or a few)
relatively hot bubble among the the mild temperature fluctuations. In
principle, a very hot bubble could lead to a runaway before the colder
ones, resulting in premature ignition in one or very few points. The
physics of the bubbles leads to a sort of self-regulation because the
efficiency of the disruption increases with the bubble
temperature (Sect.~\ref{parameter-t}) and the hottest bubbles are
disrupted more effectively. Therefore, one can 
conclude that a large range of possible ignition conditions is
unlikely.  The singly-ignited initial model proposed by \citet{cpv04}
and further analyzed by \citet{pcl04} could be interpreted as an
ignition model coming from a single-bubble runaway. This kind of
ignition is not favored by the previous probability arguments.

Of course, this homogeneity will only apply if the underlying
convective patterns in the WDs are homogeneous. The diversities in the
convective flow \citep{ww04} are potentially able to affect noticeably
the ignition conditions. Their role in producing the observed range of
diversity in SNe Ia has not yet been explored.

As far as WD central temperature, bubble temperature and radius of the
ignition zone are concerned, the results for the ignition are
essentially in agreement with the findings of \citet{wwk04} and
\citet{ww04}. While our study and the work by \citet{wwk04} and
\citet{ww04} have the same theoretical background, our approach and
methods are widely different.

We discussed the simplified assumptions that this work is based on, and their role in the analysis of the results. In particular, possible limitations are introduced by the uniform background hypothesis (Sect.~\ref{turb}) and by neglecting adiabatic cooling (Sect.~\ref{remarks}).

In SNe Ia simulations that implement multi-point ignition, the
diameter of the flame seeds (and consequently their number) is set by
the spatial resolution of the simulation. In \citet{rh04} this
diameter is $7.0\ \mbox{km}$, and future simulations, performed with
better computational resources, will be able to resolve progressively
smaller scales and allocate more bubbles. The dependence of the
explosion features on the number of bubbles, when this number is larger
than about 100, opens new insight on the explosion mechanism of SNe Ia
\citep{rhn05}. At this stage the increased spatial
resolution will not help to improve the results unless there is a
better understanding of the link between the progenitor evolution and
the early explosion phase. Future SN simulations should also take into
account the short ($\sim 0.1\ \mbox{s}$) temporal evolution of the
ignition process, and its interplay with the ongoing explosion.


\begin{acknowledgements}
The FLASH code is developed by the DOE-supported ASC / Alliance Center
for Astrophysical Thermonuclear Flashes at the University of
Chicago. L.I.~is grateful to T.~Plewa for helpful
suggestions with regards to theoretical and numerical problems, and to
the members of the FLASH Center for invaluable help during the
first FLASH Tutorial in Chicago. Thanks to S.E.~Woosley for providing the
1D models of the white dwarf used in this study, and for helpful
explanations on ignition physics. The research of J.C.N. was supported
by the Alfried Krupp Prize for Young University Teachers of the
Alfried Krupp von Bohlen und Halbach Foundation. This work was
supported in part by the European Research Training Network ``The Physics
of type Ia Supernova Explosions'' under contract HPRN-CT-2002-00303. 
\end{acknowledgements}

\bibliography{references}
\bibliographystyle{aa}

\end{document}